\documentstyle[psfrag,aps,preprint,epsfig,axodraw]{revtex}

\newcommand{\eqref}[1]{eq.(\ref{#1})}

\begin{document}

\psfrag{Hierarchical case}{\LARGE \bf Hierarchical case}
\psfrag{Degenerate case}{\LARGE \bf Degenerate case}
\psfrag{Inverted-hierarchical case}{\LARGE \bf Inverted-hierarchical case}
\psfrag{Br}{\LARGE Br}
\psfrag{Ue3}{\LARGE $U_{e3}$}
\psfrag{meg}{\LARGE $\mu \to e \gamma$}
\psfrag{m3e}{\LARGE $\mu \to eee$}
\psfrag{mec}{\LARGE $\mu - e$ conversion}
\psfrag{-11}{\Large \hspace{-5mm} $10^{-11}$}
\psfrag{-12}{\Large \hspace{-5mm} $10^{-12}~$}
\psfrag{-13}{\Large \hspace{-5mm} $10^{-13}~$}
\psfrag{-14}{\Large \hspace{-5mm} $10^{-14}~$}
\psfrag{-15}{\Large \hspace{-5mm} $10^{-15}~$}
\psfrag{-16}{\Large \hspace{-5mm} $10^{-16}~$}
\psfrag{-17}{\Large \hspace{-5mm} $10^{-17}~$}
\psfrag{-0.2}{\Large $-0.2$}
\psfrag{-0.1}{\Large $-0.1$}
\psfrag{0.0}{\Large $0$}
\psfrag{0.1}{\Large $0.1$}
\psfrag{0.2}{\Large $0.2$}

\tighten

\preprint{TU-689}
\title{Lepton flavor violation in the triplet Higgs model}
\author{Mitsuru Kakizaki\footnote{e-mail: kakizaki@tuhep.phys.tohoku.ac.jp},
Yoshiteru Ogura\footnote{e-mail: ogura@tuhep.phys.tohoku.ac.jp}
and Fumitaka Shima\footnote{e-mail: fumitaka@tuhep.phys.tohoku.ac.jp}}
\address{Department of Physics, Tohoku University,
Sendai 980-8578, Japan}
\date{\today}
\maketitle
\begin{abstract}
The triplet Higgs model, which is an extension of the standard model
with a weak-scale triplet Higgs boson,
is capable of generating small neutrino masses naturally.
We investigate lepton flavor violation mediated by the triplet Higgs boson.
We stress that various neutrino mass patterns could be distinguished 
by measuring the lepton flavor violating processes.
$\mu \to eee$ decay is significantly enhanced 
in the case of the degenerate masses or the inverted-hierarchical masses
compared with that in the case of the hierarchical masses.
On the other hand, the $\mu \to e \gamma$ rate and 
the $\mu - e$ conversion ratio in nuclei is almost insensitive 
to the mass spectra.
We also emphasize that these decay rates tend to increase as the magnitude of
$U_{e3}$ increases.
Lepton flavor nonconserving $\tau$ decay modes are 
expected to be unobservable at planned experiments
in the light of the current upper bounds of flavor violating muonic decay.
\end{abstract} 

\clearpage

Observation of neutrino oscillations and establishment of bi-large 
flavor mixing 
of the lepton sector are main progress in particle physics
in recent years.
The atmospheric neutrino experiment of the Super-Kamiokande 
implies $\nu_\mu \to \nu_\tau$ 
transition with maximal mixing\cite{Fukuda:1998mi}.
Results from the Super-Kamiokande, the SNO and the KamLAND indicate 
the large mixing angle matter-enhanced solution for solar neutrinos
\cite{Fukuda:1998fd,Ahmad:2002jz,Eguchi:2002dm}.
The small mixing angle $U_{e3}$ is required by the CHOOZ experiment
\cite{Apollonio:1999ae}.

Although the differences of the mass squared $\Delta m^2$ have been measured
by the oscillation experiments,
the absolute values of the neutrino masses remain unknown.
Direct searches of the neutrino masses such as neutrinoless double beta decay 
experiments or tritium beta decay experiments, or 
cosmological constraints cannot reach well below the eV scale.
Thus, it is important to 
seek other signals which may have some information on the 
neutrino masses.
Among other things, looking for phenomena which change generations of 
leptons is promising since 
lepton flavor violation observed in neutrino oscillations 
implies that it also occurs in the charged lepton sector.

On theoretical side,
interesting models have been proposed
that account for smallness of neutrino masses.
One representative model
is the seesaw mechanism with heavy right-handed neutrinos\cite{seesaw}.
Since the mass scale of the right-handed neutrinos is extremely high, 
phenomenological signatures but neutrino oscillations 
are negligibly suppressed.
In its supersymmetric extension with soft breaking terms, 
the situation is quite different 
since there exist new superparticles and new
flavor changing interactions at the weak scale.
Existence of the flavor off-diagonal elements of
the sfermion mass matrices and the scalar trilinear couplings
leads to observable flavor changing phenomena.
Unfortunately, these terms have
no relation to the neutrino mass matrix generically, 
so that we cannot predict anything definite without further assumptions.
An alternative to explain the neutrino masses 
is a model with an $SU(2)$ triplet Higgs field
\cite{Schechter:1980gr}.
To investigate lepton flavor violation in this model 
is particularly intriguing:
the Yukawa coupling of the triplet Higgs which generates the neutrino masses
also induces lepton flavor violating processes. 
Moreover,
the mass of the triplet Higgs can be lowered to the electroweak scale
while retaining large lepton flavor violating couplings.
Thus, new signatures, which provide us information on the neutrino masses,
could be detectable at present or future experiments.

In this letter, we explore lepton flavor violating decay
in the framework of the triplet Higgs model.
Signals of lepton flavor violation at collider experiments or at leptonic decay
experiments
in this type of models have been already discussed
\cite{Ma:2000wp,Ma:2000xh,Ma:2001mr}.
However, special forms of the mass matrices and 
specific values of the masses and the mixing angles are assumed 
in these works.
The purpose of this work is to clarify correlations 
between the lepton flavor nonconserving decay ratios and 
the mass patterns of neutrinos in a more general framework.
We will analyze muonic decay, 
say $\mu \to eee$, $\mu \to e \gamma$ and $\mu -e$ conversion in nuclei,
which gives the stringent bounds, in the three possible cases:
the hierarchical type ($m_1 \ll m_2 \ll m_3$),
the degenerate type ($m_1 \sim m_2 \sim m_3$),
and the inverted-hierarchical type ($m_3 \ll m_1 \sim m_2$).
We will show that the branching ratio of $\mu \to eee$ decay,
which arises from a tree level diagram,
depends heavily on the mass spectra while $\mu \to e \gamma$ decay and 
$\mu -e$ conversion in nuclei not.

First of all, we briefly 
review the triplet Higgs model in which the triplet Higgs possesses
a weak scale mass, concentrating on how small neutrino masses are produced
\cite{Schechter:1980gr}.
In addition to the minimal standard model fields,
an $SU(2)$ triplet scaler multiplet $\Delta$ with hypercharge 
$Y=1$ is introduced:
\begin{eqnarray}
  \begin{array}{lll}
  \Delta = \left(
    \begin{array}{cc}
      \xi^+/\sqrt{2} & \xi^{++} \\
      \xi^0 & - \xi^+/\sqrt{2} 
    \end{array}
    \right).
  \end{array}
\end{eqnarray}
The standard model gauge symmetry allows the following interaction
between the lepton doublet $l=(\nu,e)^T$
and the triplet $\Delta$ with mass $M$:
\begin{eqnarray}
  {\cal L} & = & - M^2 {\rm tr} \Delta^\dag \Delta  - \frac{1}{2} (y_N)_{ij} 
  \bar{l}^c_i \epsilon \Delta l_j + \mbox{h.c.} \nonumber \\
  & = & - M^2 (|\xi^0|^2 + |\xi^+|^2 + |\xi^{++}|^2) \nonumber \\
 & & - \frac{1}{2} (y_N)_{ij} \left[ 
    \bar{\nu}^c_i \nu_j \xi^0
    - \frac{1}{\sqrt{2}} (\bar{\nu}^c_i e_j + \bar{e}^c_i \nu_j) \xi^+
    - \bar{e}^c_i e_j \xi^{++}
    \right]  + \mbox{h.c.},
\end{eqnarray}
where $(y_N)_{ij}$ are Yukawa coupling constants and the
Latin indices $i,j$ represent generations.
Thus $\Delta$ carries lepton number $L=-2$.
After $\xi^0$ develops a vacuum expectation value,
Majorana neutrino masses which violate lepton number are generated:
\begin{eqnarray}
  m_{ij} = (y_N)_{ij} \langle \xi^0 \rangle.
\end{eqnarray}
This situation is realized
by adding the following soft lepton number violating trilinear 
interaction between
$\Delta$ and the standard model Higgs doublet $h = (h^+,h^0)^T$
to the Higgs potential:
\begin{eqnarray}
  {\cal L} & = & - \frac{1}{2}A h \epsilon
  \Delta^\dag h + \mbox{h.c.}, \nonumber \\
  & = & -\frac{1}{2} A \left[ (h^+)^2 \xi^{--} - \sqrt{2} h^+ h^0 \xi^-
    - (h^0)^2 \xi^{0\dag} \right] + \mbox{h.c.}
\end{eqnarray}
where $A$ is a mass parameter.
When $A$ is small enough compared with the weak scale, we obtain
\begin{eqnarray}
  \langle h^0 \rangle = \frac{v}{\sqrt{2}}, \quad 
  \langle \xi^0 \rangle = \frac{A v^2}{4 M^2}.
\end{eqnarray}
Here and hereafter we take universal triplet Higgs mass 
even after electroweak symmetry breakdown for simplicity.
Extension to the generic form does not alter the conclusion.
We assume that smallness of neutrino masses is attributed to tiny $A$, 
which is estimated at the eV scale in the case where 
$y_N \sim O(1)$ and $M \sim v$.
With these values, detectable lepton flavor violating processes
are expected,
whereas the constraint from the $\rho$ parameter is safely avoided.
Since lepton number is restored for $A=0$, 
it may be natural to have a small $A$ as 
a consequence of tiny lepton number violation.
Smallness of the lepton flavor violating interaction 
can be explained in the context of large extra dimensions
\cite{Ma:2000wp,Ma:2000xh}.
We will not discuss other possible origins of this interaction in this paper.

We now concentrate on flavor structure of this model.
One-to-one correspondence between the neutrino mass matrix and $(y_N)_{ij}$
is of particular importance:
\begin{eqnarray}
  m_{ij} = \frac{A v^2}{4 M^2} (y_N)_{ij}.
\end{eqnarray}
Hence, we can distinguish neutrino mass patterns by measuring $(y_N)_{ij}$.
As mentioned before, there exist three possible types of neutrino mass spectra.
In terms of $\Delta m^2_{\rm sol}$ and  $\Delta m^2_{\rm atm}$
obtained by the experiments, they are classified into 
\begin{itemize}
\item Hierarchical type ($m_1 \ll m_2 \ll m_3$):
  \begin{eqnarray}
    m_1 = 0, \quad 
    m_2 = \sqrt{\Delta m^2_{\rm sol}}, \quad
    m_3 = \sqrt{\Delta m^2_{\rm atm}}.
  \end{eqnarray}
\item Degenerate type ($m_1 \sim m_2 \sim m_3$):
  \begin{eqnarray}
    m_1 = m_\nu, \quad 
    m_2 = m_\nu + \frac{\Delta m^2_{\rm sol}}{2 m_\nu}, \quad 
    m_3 = m_\nu + \frac{\Delta m^2_{\rm atm}}{2 m_\nu}.
  \end{eqnarray}
\item Inverted-hierarchical type ($m_3 \ll m_1 \sim m_2$):
  \begin{eqnarray}
    m_1 = m_2 - \frac{\Delta m^2_{\rm sol}}{2 m_2}, \quad 
    m_2 = \sqrt{\Delta m^2_{\rm atm}}, \quad
    m_3 = 0.
  \end{eqnarray}
\end{itemize}
In terms of the standard parametrization, 
the unitary matrix which diagonalize the neutrino mass matrix $m_{ij}$
is given by 
\begin{eqnarray}
  U & = & VP, \nonumber \\
  V & = & \left(
    \begin{array}{ccc}
      c_{13}c_{12} & c_{13}s_{12} & s_{13}e^{-i\phi} \\
      - c_{23}s_{12} - s_{23}s_{13}c_{12}e^{i\phi} & 
      c_{23}c_{12} - s_{23}s_{13}s_{12}e^{i\phi} & s_{23}c_{13} \\
      s_{23}s_{12} - c_{23}s_{13}c_{12}e^{i\phi} & 
      - s_{23}c_{12} - c_{23}s_{13}s_{12}e^{i\phi} & c_{23}c_{13}
    \end{array}
  \right), \nonumber \\
  P & = & \left(
    \begin{array}{ccc}
      1 & 0 & 0 \\
      0 & e^{i\phi_2} & 0 \\
      0 & 0 & e^{i\phi_3} 
    \end{array}
    \right)
\end{eqnarray}
where $s_{ij} \equiv \sin \theta_{ij}$ and $c_{ij} \equiv \cos \theta_{ij}$,
$\phi$ stands for the Dirac phase and $\phi_2,\phi_3$ the Majorana ones which 
are responsible for $CP$ violation.
For simplicity, 
we take the following typical values in our later 
evaluation of the lepton flavor
violating processes\cite{Pakvasa:2003zv}:
\begin{eqnarray}
  \Delta m^2_{\rm sol} = 7.0 \times 10^{-5} ~\mbox{eV}^2, \quad 
  \Delta m^2_{\rm atm} = 2.5 \times 10^{-3} ~\mbox{eV}^2,
\end{eqnarray}
\begin{eqnarray}
  s_{12} = 0.5, \quad s_{23} = 1/\sqrt{2}
\end{eqnarray}
which are favored by the solar and the atmospheric neutrino data respectively 
and the bound $s_{13} \leq 0.2$ from the reactor experiment CHOOZ,
and we ignore possible $CP$ violating phases.

We are now at a position to consider dependence of lepton flavor violation 
on the neutrino mass patterns in this model.
Large mixing in the lepton mixing matrix and presence of the weak-scale 
triplet Higgs give rise to  dangerous lepton flavor violating processes.
The severest bounds come from muonic decay modes
such as $\mu \to eee, \mu \to e \gamma$ and $\mu - e$ conversion
in nuclei.
The present upper bounds of these modes are 
$\mbox{Br} (\mu \to e \gamma) < 1.2 \times 10^{-11}$\cite{Brooks:1999pu}, 
$\mbox{Br} (\mu \to eee) < 1.0 \times 10^{-12}$\cite{Bellgardt:1987du}
and $R(\mu \mbox{Ti} \to e \mbox{Ti}) < 4.3 \times 10^{-12}$ \cite{Dohmen:mp}.
Future experiments will reach 
$\mbox{Br} (\mu \to e \gamma) \sim 10^{-14}$ \cite{psi} and 
$R(\mu \mbox{Al} \to e \mbox{Al}) \sim 10^{-16}$ \cite{meco} .

We give the formulae for these processes in the triplet Higgs model.
The $\mu \rightarrow eee$ process occurs at tree level in this model and
the branching ratio is calculated as 
\begin{eqnarray}
  \mbox{Br}(\mu \to eee) 
  = \frac{1}{64 G_F^2} 
  \frac{|(y_N^\dag)_{11}(y_N)_{21}|^2}{M^4}
\end{eqnarray}
where $G_F$ is the Fermi coupling constant.
The off-shell amplitude of $\mu \rightarrow e \gamma$ can be written
\begin{eqnarray}
  {\cal M} = e \epsilon^\alpha j_\alpha
\end{eqnarray}
where $e$ is the electric charge, $\epsilon^\alpha$ is the photon polarization,
and $j_\alpha$ is the leptonic current given by
\begin{eqnarray}
  j_\alpha = \bar{u}_e(p+q)[q^2 \gamma_\alpha (A_1^L P_L + A_1^R P_R)
  + m_\mu i \sigma_{\alpha \beta} q^\beta (A_2^L P_L + A_2^R P_R)]u_\mu(p).
\end{eqnarray}
for small momentum transfer $q$.
Here $P_{L,R} = (1 \mp \gamma^5)/2$.
The branching ratio of $\mu \to e\gamma$ is calculated as
\begin{eqnarray}
  \mbox{Br} (\mu \rightarrow e \gamma)
  & = & \frac{48 \pi^3 \alpha}{G_F^2}(|A_2^L|^2 + |A_2^R|^2).
\end{eqnarray}
The matrix element of photon exchange which contribute to $\mu - e$ conversion
in nuclei is written
\begin{eqnarray}
  {\cal M} = \frac{e^2}{q^2} j^\alpha J_\alpha
\end{eqnarray}
where $J_\alpha$ is the hadronic current.
The coherent $\mu - e$ conversion ratio is calculated as 
\begin{eqnarray}
  R = \frac{4 \alpha^5 m_\mu^5 Z_{\rm eff}^4 Z |F(q)|^2}{\Gamma_{\rm capt}}
    (|A_1^L + A_2^R|^2 + |A_1^R + A_2^L|^2).
\end{eqnarray}
For $^{48}_{22} \mbox{Ti}$, 
$Z_{\rm eff} = 17.6, F(q^2 \approx - m_\mu^2) \approx 0.54$ 
and $\Gamma_{\rm capt} \approx 2.6 \times 10^6~ \mbox{s}^{-1}$
\cite{Bernabeu:1993ta}.
The form factors are produced by one-loop diagrams
of the triplet Higgs exchange:
\begin{eqnarray}
  A_1^L & = & 
  (y_N^\dag)_{1k}(y_N)_{k2} \frac{1}{16\pi^2} \frac{1}{3 M^2} 
  F(r,s_k) \nonumber \\
  A_1^R & = & 0 \nonumber \\
  A_2^L & = & 0 \nonumber \\
  A_2^R & = & - (y_N^\dag y_N)_{12} \frac{1}{16\pi^2} \frac{3}{8 M^2}, 
  \nonumber \\
  F(r,s_k) & \equiv & \ln s_k + \frac{4 s_k}{r} +  
  \left( 1 -\frac{2s_k}{r} \right) \sqrt{1 + \frac{4s_k}{r}} 
  \ln \frac{\sqrt{r + 4 s_k} + \sqrt{r}}{\sqrt{r + 4 s_k} - \sqrt{r}}
  \label{eq:f}
\end{eqnarray}
where $r \equiv - q^2/M^2, s_k \equiv m_k^2/M^2$.

Here we would like to emphasize characteristics of the triplet Higgs model:
\begin{itemize}
\item 
The $\mu \to eee$ process occurs at tree level while $\mu \to e \gamma$ and 
$\mu - e$ conversion at one-loop level.
Thus, the $\mu \to eee$ rate tends to be larger 
compared to the case where $\mu \to eee$ arises at one-loop level, 
like in the supersymmetric standard model.

\item 
Since the triplet Higgs couples only to the left-handed leptons,
$A_2^L$ vanishes. 
This situation is similar to the minimal supersymmetric standard model 
with right-handed neutrinos where $A_2^R$ dominates over $A_2^L$
\cite{Borzumati:1986qx,Hisano:1995nq}.
On the other hand, in the minimal $SU(5)$ supersymmetric grand unified theory,
$A_2^L$ is dominant\cite{Barbieri:1994pv}.
Whether $A_2^R$ dominates or not may be tested 
by measuring the angular distribution of $e^+$s 
in $\mu^+ \to e^+ \gamma$ decay if polarized muon is available
\cite{Kuno:1996kv}.

\item
In models where $A_1^L \sim A_2^R$ is realized, 
$R/\mbox{Br}(\mu \to e \gamma) \sim O(10^{-2})$ would be predicted.
On the other hand, in this model $A_1^L$ is enhanced 
by $\log (m_\mu^2/M^2)$ compared to $A_2^R$ (See \eqref{eq:f}) ,
so that the ratio $R/\mbox{Br}(\mu \to e \gamma)$ is not so suppressed
\cite{Marciano:cj,Raidal:1997hq}.
In fact, the $\mu - e$ conversion ratio is comparable to the 
the $\mu \to e \gamma$ branching fraction, as we will see below.

\item 
Since the neutrino masses are fixed, $y_N$ is proportional to $M^2$ and 
thus the lepton flavor violating rates to $M^4$.
Therefore, searches for lepton flavor violating decay and 
collider experiments are complementary to each other.

\end{itemize}

Let us discuss dependence of these ratios on the mass patterns qualitatively
before presenting numerical calculation.
In terms of the observed values,
$(m^\dag m)_{12}$, which is included in the decay rate of $\mu \to e \gamma$,
is expressed as, 
\begin{eqnarray}
  (m^\dag m)_{12} \sim \left\{
    \begin{array}{lll}
      \frac{\sqrt{6}}{8} \Delta m^2_{\rm sol}
      + \frac{\sqrt{2}}{2} \Delta m^2_{\rm atm} s_{13}.
      \quad (\mbox{Hierarchical type}) \\
      \frac{\sqrt{6}}{8} \Delta m^2_{\rm sol}
      + \frac{\sqrt{2}}{2} \Delta m^2_{\rm atm} s_{13}.
      \quad (\mbox{Degenerate type}) \\
      \frac{\sqrt{6}}{8} \Delta m^2_{\rm sol}
      - \frac{\sqrt{2}}{2} \Delta m^2_{\rm atm} s_{13}.
      \quad (\mbox{Inverted-hierarchical type})
    \end{array}
  \right. .
  \label{eq:meg}
\end{eqnarray}
This implies that the $\mu \to e \gamma$ branching fraction is 
almost independent of the mass patterns. 
On the contrary, $m_{11}^\dag m_{12}$, which is involved in 
the $\mu \to eee$ rate, is divided into three patterns:
\begin{eqnarray}
  m_{11}^\dag m_{12} \sim \left\{
    \begin{array}{lll}
      \frac{\sqrt{6}}{32} \Delta m^2_{\rm sol}
      + \frac{\sqrt{2}}{8}\sqrt{\Delta m^2_{\rm sol} \Delta^2_{\rm atm}}s_{13} 
      \quad (\mbox{Hierarchical type}) \\
      \frac{\sqrt{6}}{16} \Delta m^2_{\rm sol} 
      + \frac{\sqrt{2}}{4} \Delta m^2_{\rm atm} s_{13}
      \quad (\mbox{Degenerate type}) \\
      \frac{\sqrt{6}}{16} \Delta m^2_{\rm sol} 
      - \frac{\sqrt{2}}{2} \Delta m^2_{\rm atm} s_{13} 
      \quad (\mbox{Inverted-hierarchical type})
    \end{array}
  \right. .
  \label{eq:m3e}
\end{eqnarray}
This behavior indicates that the $\mu \to eee$ rate is very sensitive to
the neutrino mass structure.
The expression is complicated for the $\mu - e$ conversion ratio,
and we defer this point to later numerical estimation.
For relatively large $s_{13}$, 
one finds that the first terms in \eqref{eq:meg} and 
\eqref{eq:m3e} are negligible and 
that the $\mu \to eee$ ratio becomes substantially enhanced
in the degenerate or the inverted-hierarchical types.

\begin{figure}[ht]
  \begin{center}
    \makebox[0cm]{
      \scalebox{0.55}{
        \includegraphics{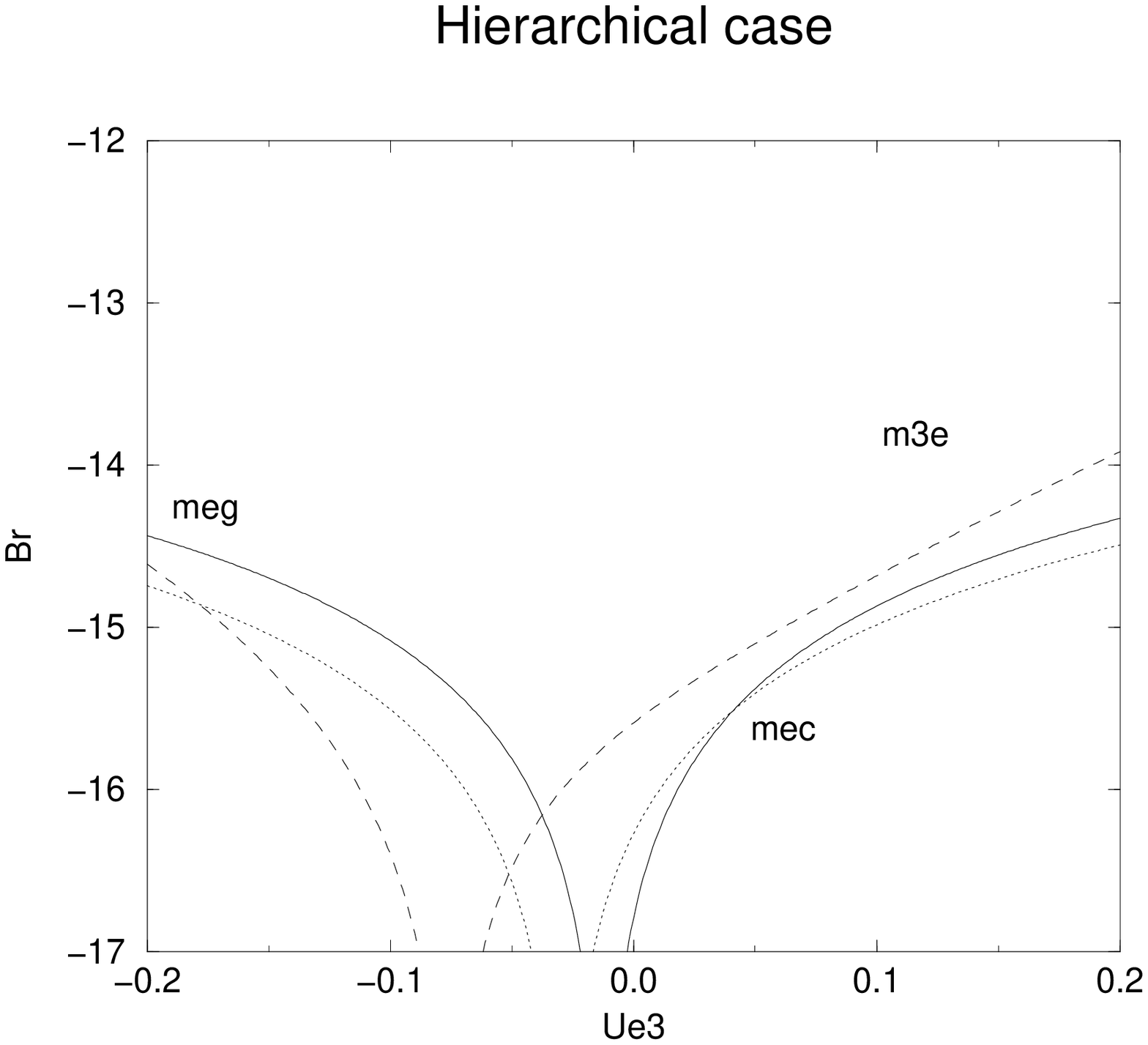}
        }
      }
    \caption{The branching ratios of the processes $\mu \to e \gamma$ (the solid line), $\mu \to eee$ (the dashed line), $\mu$ - $e$ conversion in Ti (the dotted line) for the hierarchical case. Here $M=200~\mbox{GeV}$,$A=25~\mbox{eV}$ are taken. }
    \label{fig:hierarchy}
  \end{center}
\end{figure}

\begin{figure}[ht]
  \begin{center}
    \makebox[0cm]{
      \scalebox{0.55}{
        \includegraphics{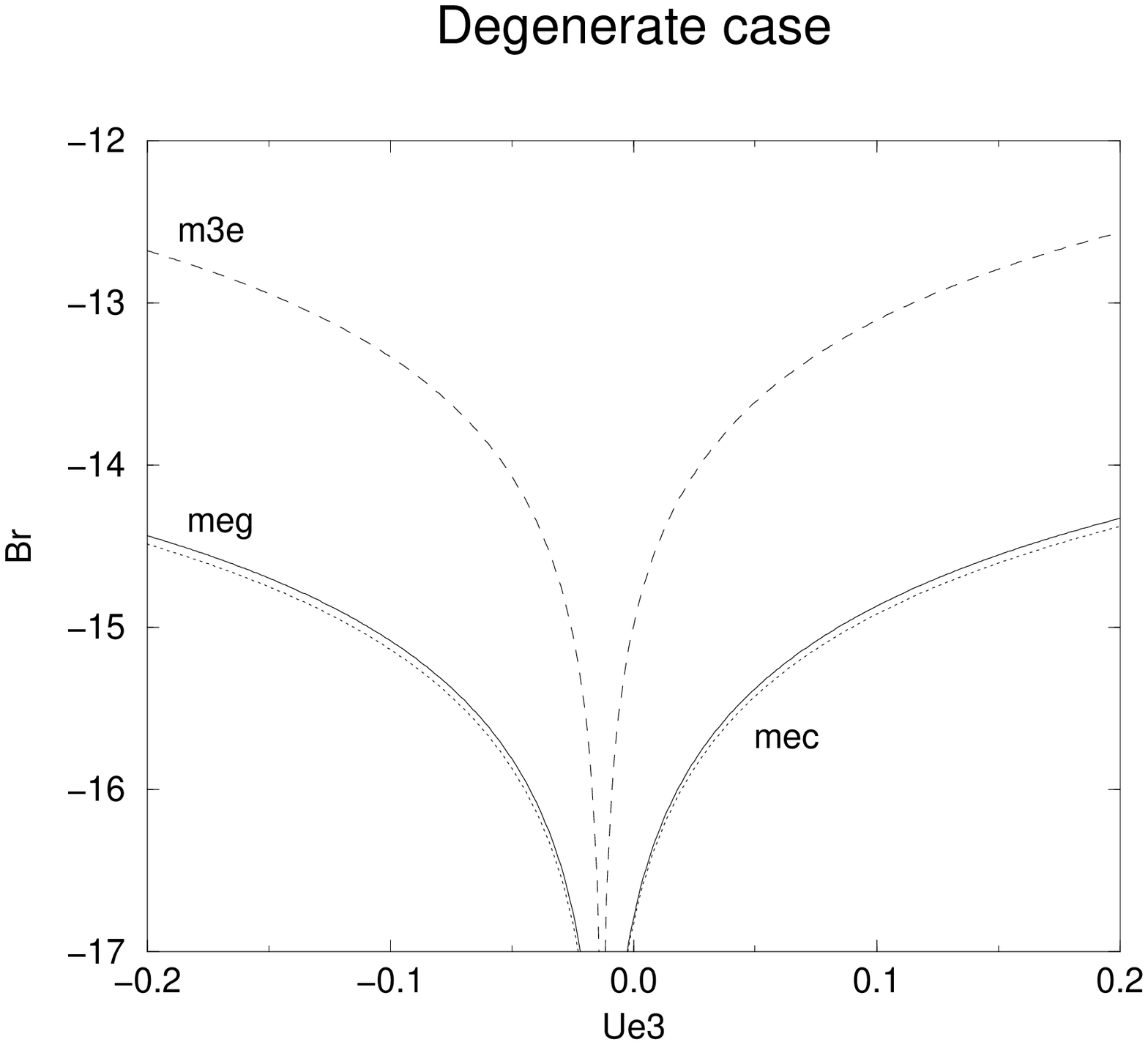}
        }
      }
    \caption{The branching ratios of the processes $\mu \to e \gamma$ (the solid line), $\mu \to eee$ (the dashed line), $\mu$ - $e$ conversion in Ti (the dotted line) for the degenerate case. Here $m_\nu = 0.1 ~\mbox{eV}$ is taken. The other parameters are same as Fig. \ref{fig:hierarchy}.} 
    \label{fig:degeneracy}
  \end{center}
\end{figure}

\begin{figure}[ht]
  \begin{center}
    \makebox[0cm]{
      \scalebox{0.55}{
        \includegraphics{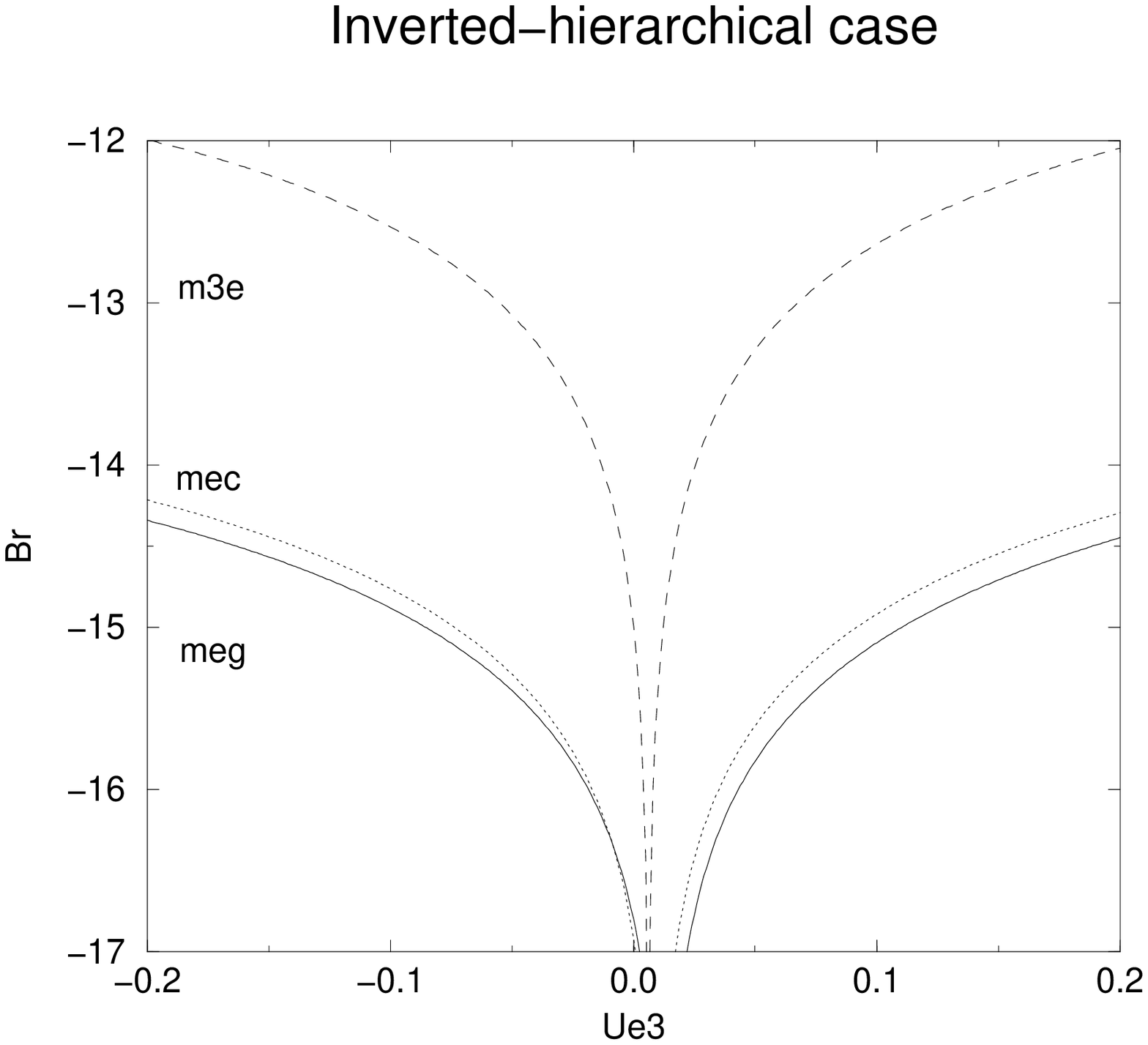}
        }
      }
    \caption{The branching ratios of the processes $\mu \to e \gamma$ (the solid line), $\mu \to eee$ (the dashed line), $\mu$ - $e$ conversion in Ti (the dotted line) for the inverted-hierarachical case. The other parameters are same as Fig. \ref{fig:hierarchy}.}
    \label{fig:inverse}
  \end{center}
\end{figure}

We have performed numerical calculation of the ratios,
elucidating mass and mixing angle dependence.
The ratios as a function of $U_{e3}$ are plotted in Figs. \ref{fig:hierarchy} 
- \ref{fig:inverse} for three types of the mass spectra.
Here the solid line, the dashed one and the dotted one 
correspond to $\mbox{Br}(\mu \to e \gamma)$, 
$\mbox{Br}(\mu \to eee)$ and $R(\mu \mbox{Ti} \to e\mbox{Ti})$ respectively.
We take the triplet Higgs mass at $M=200~\mbox{GeV}$ and
the trilinear coupling between the Higgs multiplets at $A=25~\mbox{eV}$.
The relative ratios among these decay rates remain unchanged 
for arbitrary $M$ or $A$
since all of the absolute values of the three ratios are proportional 
to $M^4/A^4$ 
(though $\mu - e$ conversion slightly depends on $M$ itself. 
See \eqref{eq:f}.).
Our main concern is the relative ratios but not the absolute values which 
are theoretically unrestricted in our setup.
For the degenerate case (Fig. \ref{fig:degeneracy}), 
the universal neutrino mass is fixed at $m_\nu = 0.1 \mbox{eV}$ 
in order not to conflict the constraint from the resent WMAP data,
$m_\nu \lesssim 0.23 ~\mbox{eV}$ \cite{Spergel:2003cb}.
One can verify that $\mu \to eee$ is strongly related to
the mass eigenvalues while $\mu \to e \gamma$ and $\mu - e$ conversion in 
nuclei not.
Roughly speaking, the ratios are proportional to $s_{13}^2$
as one understands from \eqref{eq:meg} and \eqref{eq:m3e}.
One of the most interesting points is the fact that, as expected,
the $\mu \to eee$ ratio dominates over 
the $\mu \to e \gamma$ one and the $\mu - e$ conversion ratio
in the degenerate and the inverted-hierarchical cases.
In the case of the hierarchical masses,
the three ratios are almost equal for positive $U_{e3}$,
and there is a region where the 
$\mu \to e \gamma$ ratio dominates for negative $U_{e3}$
because cancellation among terms occurs for the other processes. 
Thus, we conclude that the neutrino mass pattern
could be determined once we measure the muonic decay modes which convert the
generations of the leptons.

Finally, we would like to mention flavor violating $\tau$ decay.
The $\tau$ decay modes, such as $\tau \to \mu \gamma$ 
and $\tau \to \mu \mu \mu$, 
are not so enhanced compared with the muonic ones in this framework
because of the large mixing angles in the lepton sector.
Since muonic decay already puts the severe bound
$\mbox{Br}(\mu \to eee) < 10^{-12}$,
$\tau$ decay cannot be observed even at the next generation experiments,
which aim at $\mbox{Br}(\tau \to \mu \gamma) \sim 10^{-7}$--$10^{-8}$
\cite{tmg},
in the absence of accidental cancellation in the muonic decay rates.

In conclusion, 
searches for lepton flavor violation are crucial
not only to confirm existence of the triplet Higgs which 
accounts for the neutrino masses but also 
to determine the absolute values of the neutrinos in this framework.
We investigate how lepton flavor violating decay 
relies on the neutrino mass spectra in the triplet Higgs model.
We show that the $\mu \to eee$ ratio is largely enhanced compared with 
the $\mu \to e \gamma$ ratio and the $\mu - e$ conversion ratio
in the case of the degenerate neutrino mass spectra 
and the inverted-hierarchical ones.
On the other hand, 
the three ratios are equal generically in the hierarchical masses.
As for mixing angle dependence, 
the decay rates incline to be enhanced as $|U_{e3}|$ increases.
The muon can decay only through $\mu^+ \to e^+_R \gamma$,
which might be examined
by observing angular distribution of $e^+$s
if future experiments with polarized muon are available.
Flavor violating $\tau$ decay would not be observed in the near future
if the neutrino-mass-giving Yukawa interactions are only sources of 
lepton flavor violation.
These characteristic signatures should be compared with predictions of 
other models which cause observable lepton flavor violation,
like supersymmetric models.

\

 After completion of this work, we received a preprint\cite{Chun:2003ej}
which deals with a similar subject.

\section*{Acknowledgment}               
The authors would like to thank M. Yamaguchi for 
valuable discussion and careful reading of the manuscript,
T. Moroi for suggesting this interesting subject and valuable discussion, 
and N. Abe and M. Endo for useful discussion.


\begin{references}


\bibitem{Fukuda:1998mi}
Y.~Fukuda {\it et al.}  [Super-Kamiokande Collaboration],
Phys.\ Rev.\ Lett.\  {\bf 81} (1998) 1562
[arXiv:hep-ex/9807003].



\bibitem{Fukuda:1998fd}
Y.~Fukuda {\it et al.}  [Super-Kamiokande Collaboration],
Phys.\ Rev.\ Lett.\  {\bf 81} (1998) 1158
[Erratum-ibid.\  {\bf 81} (1998) 4279]
[arXiv:hep-ex/9805021];

S.~Fukuda {\it et al.}  [Super-Kamiokande Collaboration],
Phys.\ Rev.\ Lett.\  {\bf 86} (2001) 5656
[arXiv:hep-ex/0103033].

\bibitem{Ahmad:2002jz}
Q.~R.~Ahmad {\it et al.}  [SNO Collaboration],
Phys.\ Rev.\ Lett.\  {\bf 89} (2002) 011301
[arXiv:nucl-ex/0204008].


\bibitem{Eguchi:2002dm}
K.~Eguchi {\it et al.}  [KamLAND Collaboration],
Phys.\ Rev.\ Lett.\  {\bf 90} (2003) 021802
[arXiv:hep-ex/0212021].



\bibitem{Apollonio:1999ae}
M.~Apollonio {\it et al.}  [CHOOZ Collaboration],
Phys.\ Lett.\ B {\bf 466} (1999) 415
[Erratum-ibid.\ B {\bf 472} (2000) 434]
[arXiv:hep-ex/9907037].



\bibitem{seesaw} 
T.~Yanagida, in Proc. of the
workshop on the Unified Theory and Baryon Number in the Universe,
eds. O.~Sawada and A.~Sugamoto (KEK, Tsukuba, 1979) 95;

M.~Gell-Mann, P.~Ramond, R.~Slansky, in Supergravity,
ed. by P.~van Nieuwenhuizen and D.~Freedman (Amsterdam,
North Holland, 1979) 315.



\bibitem{Schechter:1980gr}
J.~Schechter and J.~W.~Valle,
Phys.\ Rev.\ D {\bf 22} (1980) 2227.



\bibitem{Ma:2000wp}
E.~Ma, M.~Raidal and U.~Sarkar,
Phys.\ Rev.\ Lett.\  {\bf 85} (2000) 3769
[arXiv:hep-ph/0006046].



\bibitem{Ma:2000xh}
E.~Ma, M.~Raidal and U.~Sarkar,
Nucl.\ Phys.\ B {\bf 615} (2001) 313
[arXiv:hep-ph/0012101].



\bibitem{Ma:2001mr}
E.~Ma and M.~Raidal,
Phys.\ Rev.\ Lett.\  {\bf 87} (2001) 011802
[Erratum-ibid.\  {\bf 87} (2001) 159901]
[arXiv:hep-ph/0102255].



\bibitem{Pakvasa:2003zv}
S.~Pakvasa and J.~W.~Valle,
arXiv:hep-ph/0301061.



\bibitem{Brooks:1999pu}
M.~L.~Brooks {\it et al.}  [MEGA Collaboration],
Phys.\ Rev.\ Lett.\  {\bf 83} (1999) 1521
[arXiv:hep-ex/9905013].



\bibitem{Bellgardt:1987du}
U.~Bellgardt {\it et al.}  [SINDRUM Collaboration],
Nucl.\ Phys.\ B {\bf 299} (1988) 1.




\bibitem{Dohmen:mp}
C.~Dohmen {\it et al.}  [SINDRUM II Collaboration.],
Phys.\ Lett.\ B {\bf 317} (1993) 631.


\bibitem{psi}
D.~Nicolo', talk at the workshop ``Neutrino oscillations and their origin'' (NOON2003) (Kanazawa, Japan, Feb.,2003)


\bibitem{meco}
A. Van der Schaaf, talk at the workshop ``Neutrino oscillations and their origin'' (NOON2003) (Kanazawa, Japan, Feb.,2003)



\bibitem{Bernabeu:1993ta}
J.~Bernabeu, E.~Nardi and D.~Tommasini,
Nucl.\ Phys.\ B {\bf 409} (1993) 69
[arXiv:hep-ph/9306251].



\bibitem{Borzumati:1986qx}
F.~Borzumati and A.~Masiero,
Phys.\ Rev.\ Lett.\  {\bf 57} (1986) 961.



\bibitem{Hisano:1995nq}
J.~Hisano, T.~Moroi, K.~Tobe, M.~Yamaguchi and T.~Yanagida,
Phys.\ Lett.\ B {\bf 357} (1995) 579
[arXiv:hep-ph/9501407];

J.~Hisano, T.~Moroi, K.~Tobe and M.~Yamaguchi,
Phys.\ Rev.\ D {\bf 53} (1996) 2442
[arXiv:hep-ph/9510309].


\bibitem{Barbieri:1994pv}
R.~Barbieri and L.~J.~Hall,
Phys.\ Lett.\ B {\bf 338} (1994) 212
[arXiv:hep-ph/9408406].



\bibitem{Kuno:1996kv}
Y.~Kuno and Y.~Okada,
Phys.\ Rev.\ Lett.\  {\bf 77} (1996) 434
[arXiv:hep-ph/9604296].



\bibitem{Marciano:cj}
W.~J.~Marciano and A.~I.~Sanda,
Phys.\ Rev.\ Lett.\  {\bf 38} (1977) 1512.



\bibitem{Raidal:1997hq}
M.~Raidal and A.~Santamaria,
Phys.\ Lett.\ B {\bf 421} (1998) 250
[arXiv:hep-ph/9710389].



\bibitem{Spergel:2003cb}
D.~N.~Spergel {\it et al.},
arXiv:astro-ph/0302209.



\bibitem{tmg}
T.~Ohshima, talk at the workshop ``Neutrino oscillations and their origin'' (NOON2001) (ICRR, Univ. of Tokyo, Kashiwa, Japan, Dec.,2001)





\bibitem{Chun:2003ej}
E.~J.~Chun, K.~Y.~Lee and S.~C.~Park,
arXiv:hep-ph/0304069.


\end{references}
\end{document}